\documentclass{PoS}
\usepackage{amsmath,amsthm,amssymb,amsfonts}

\title{Dynamical parton distributions and weak-gauge and Higgs boson production at hadron colliders at NNLO of QCD}

\ShortTitle{Dynamical parton distributions and weak-gauge and Higgs boson ...}

\author{\speaker{P. Jimenez-Delgado}\thanks{Supported by the Swiss National Science Foundation (SNF) under contract 200020-126691.}\\University of Zurich\\E-mail: \email{pjimenez@physik.uzh.ch}}

\author{E. Reya\thanks{Supported in part by the ``Bundesministerium f\"ur Bildung und Forschung'', Berlin.}\\TU Dortmund\\E-mail: \email{reya@physik.tu-dortmund.de}}

\FullConference{XVIII International Workshop on Deep-Inelastic Scattering and Related Subjects\\April 19 -23, 2010\\Convitto della Calza, Firenze, Italy}

\abstract{Utilizing recent DIS measurements and data on hadronic dilepton production we determine at NNLO (3-loop) of QCD the dynamical parton distributions of the nucleon generated radiatively from valencelike positive input distributions at an optimally chosen low resolution scale ($Q_0^2\!<\!1$~GeV$^2$) by employing the ``fixed flavor number factorization scheme'' (FFNS). These are compared with ``standard'' NNLO distributions, generated at some fixed and higher resolution scale ($Q_0^2\!>\!1$~GeV$^2$). The NNLO corrections imply in both approaches an improved value of $\chi^2$, typically $\chi^2_{NNLO}\sim0.9\chi^2_{NLO}$. The dynamical NNLO uncertainties are somewhat smaller than the NLO ones and both are, as expected, smaller than those of their ``standard'' counterparts. The dynamical predictions for $F_L(x,Q^2)$ become perturbatively stable already at $Q^2 = 2$ - 3~GeV$^2$, where precision measurements could even delineate NNLO effects in the very small-x region. We obtain $\alpha_s(M_Z^2) =$ 0.1124 $\pm$ 0.0020 to be compared with 0.1145 $\pm$ 0.0018 at NLO.

Using these NNLO dynamical parton distributions in the FFNS as input, we generate radiatively parton distributions in the ``variable flavor number factorization scheme'' (VFNS) as well, where also the heavy quark flavors (c,b,t) become massless partons within the nucleon. Only within the VFNS are NNLO calculations feasible at present, since the required partonic subprocesses are available only in the approximation of massless initial-state partons. The NNLO predictions for gauge boson production are typically larger (by more than 1$\sigma$) than the NLO ones, and rates at LHC energies can be predicted with an accuracy of about 5\%, whereas at Tevatron they are more than 2$\sigma$ above the NLO ones. The NNLO predictions for SM Higgs boson production via the dominant gluon fusion process have a total (PDFs and scale) uncertainty of about 10\% at LHC which almost doubles at the lower Tevatron energies; these predictions are typically about 20\% larger than the ones at NLO but the total uncertainty bands overlap.}

\begin{document}

The \emph{dynamical} parton distributions of the nucleon at $Q^2\!\gtrsim\!1$~GeV$^2$ are QCD radiatively generated from {\em valencelike}\footnote{Valencelike refers to $a_f\!>\!0$ for {\em all} input distributions $xf(x,Q_0^2)\propto x^{a_f}(1-x)^{b_f}$, i.e., not only the valence but also the sea and gluon input densities vanish at small $x$.} positive definite input distributions at an optimally determined low input scale $Q_0^2\!<\!1$~GeV$^2$. Therefore the \emph{steep} small-Bjorken-$x$ behavior of structure functions, and consequently of the gluon and sea distributions, appears within the dynamical (radiative) approach mainly as a consequence of QCD-dynamics at $x \lesssim 10^{-2}$ \cite{Gluck:1994uf}. Alternatively, in the common ``standard'' approach the input scale is fixed at some arbitrarily chosen $Q^2_0\!>\!1$~GeV$^2$, and the corresponding input distributions are less restricted; for example, the mentioned {\em steep} small-$x$ behavior has to be {\em fitted}.

Following the radiative approach, the well-known LO/NLO GRV98 dynamical parton distribution functions of \cite{Gluck:1998xa} have been updated in \cite{Gluck:2007ck}, and the analysis extended to the NNLO of perturbative QCD in \cite{JimenezDelgado:2008hf}. In addition, in \cite{Gluck:2007ck, JimenezDelgado:2008hf} a series of ``standard'' fits were produced in (for the rest) exactly the same conditions as their dynamical counterparts. This allows us to compare the features of both approaches and to test the the dependence in model assumptions. The associated uncertainties encountered in the determination of the parton distributions turn out, as expected, to be larger in the ``standard'' case, particularly in the small-$x$ region, than in the more restricted dynamical radiative approach where, moreover, the ``evolution distance'' (starting at $Q_0^2\!<\!1$~GeV$^2$) is sizably larger \cite{Gluck:2007ck, JimenezDelgado:2008hf}.

The NNLO corrections imply in both approaches an improved value of $\chi^2$, typically $\chi^2_{NNLO}\sim0.9\chi^2_{NLO}$. The dynamical NNLO uncertainties are somewhat smaller than the NLO ones and both are smaller than those of their ``standard'' counterparts. The strong coupling constant $\alpha_s(M_Z^2)$ is determined in our analyses together with the parton distributions, in particular it is closely related to the gluon distribution which drives the QCD evolution and consequently its uncertainty is also smaller in the dynamical case. We obtain $\alpha_s(M_Z^2) =$ 0.1124 $\pm$ 0.0020 at NNLO, and 0.1145 $\pm$ 0.0018 at NLO in the dynamical case; to be compared with $\alpha_s(M_Z^2) =$ 0.1158 $\pm$ 0.0035 at NNLO, and 0.1178 $\pm$ 0.0021 at NLO in the ``standard'' one\footnote{See \cite{JimenezDelgado:2008hf} for a more detailed discussion.}. The dynamical predictions for $F_L(x,Q^2)$ become perturbatively stable already at $Q^2\!=\!2$ - 3~GeV$^2$, where precision measurements could even delineate NNLO effects in the very small-x region. This is in contrast to the results in the common ``standard'' approach, but NNLO/NLO differences are there less distinguishable due to the larger uncertainty bands$^2$.

With the LHC and Tevatron running and having in mind that parton distributions are one of the major sources of uncertainty in the predictions at hadron colliders, we will focus in this talk on the implications of our NNLO distributions, and specially of their uncertainties, for important process like weak gauge boson production and the production of the standard model (SM) Higgs boson itself. These results have been published in \cite{JimenezDelgado:2009tv}, where more details and further necessary references have been given. 

The analyses in \cite{Gluck:2007ck, JimenezDelgado:2008hf} were performed within the framework of the so-called  ``fixed flavor number scheme'' (FFNS) where, besides the gluon, only the light quark flavors $q=u,d,s$ are considered as genuine, i.e., massless partons within the nucleon. This factorization scheme is fully predictive in the heavy quark $h=c,b,t$ sector where the heavy quark flavors are produced entirely perturbatively as part of the final state. Here the full heavy quark mass $m_h$ dependence is taken into account in the production cross sections, as required experimentally, in particular, in the threshold region. Even for very large values of $Q^2\gg m_{c,b}^2$, the FFNS predictions are in remarkable agreement with deep inelastic scattering data and, moreover, are perturbatively stable despite the common belief that ``non-collinear'' logarithms $\ln\tfrac{Q^2}{m_h^2}$ have to be resummed. This agreement with experiment even at $Q^2\gg m_h^2$ indicates that there is {\em little need} to resum these supposedly ``large logarithms'', which is of course in contrast to the genuine collinear logarithms appearing in light (massless) quark and gluon hard scattering processes.

However, in many situations calculations within the FFNS become unduly complicated, thus it is of practical advantage to consider the so-called ``variable flavor number scheme'' (VFNS) in which the heavy quarks are considered to be (massless) partons within the nucleon as well. This factorization scheme is characterized by increasing the number of flavors $n_f$ of massless partons by one unit at $Q^2\!=\!m_h^2$ starting from $n_f\!=\!3$ at $Q^2=m_c^2$.  Hence the $n_f\!>\!3$ ``heavy'' quark distributions are perturbatively uniquely generated from the $n_f-1$ ones via the massless renormalization group $Q^2$-evolution; a comparative qualitative and quantitative discussion of this (zero-mass) VFNS and the FFNS has been recently presented in \cite{Gluck:2008gs}. Eventually one has to {\em assume} that these massless ``heavy'' quark distributions are relevant asymptotically, i.e., that they correctly describe the asymptotic behavior of DIS structure functions for scales $Q^2\gg m_h^2$.  However, for most experimentally accessible values of $Q^2$, in particular around the threshold region of heavy quark $(h\bar{h})$ production, effects due to {\em finite} heavy quark masses $m_h$ can {\em not} be neglected. One therefore needs either to stick to the FFNS or an improvement of this zero-mass VFNS which maintain heavy quark mass-dependent corrections in the hard cross-sections and interpolate between the zero-mass VFNS (assumed to be correct asymptotically) and the (experimentally required for most data) FFNS. Such improvements are often referred to as general-mass VFNS and there exist various different model-dependent ways of implementing the required $m_h$ dependence (see, e.g. \cite{Gluck:2008gs} for references).

In order to avoid any such model ambiguities we generate ``heavy''-quark zero-mass VFNS distributions using our unique NNLO dynamical FFNS distributions as input at $Q^2\!=\!m_c^2$. This considerably eases the otherwise unduly complicated calculations in the FFNS of weak gauge- and Higgs-boson production at hadron collider energies. It has been shown \cite{Gluck:2008gs} that for situations where the invariant mass of the produced system ($cW,\, tW,\, t\bar{b},\,$ Higgs-bosons, etc.) exceeds by far the mass of the participating heavy flavor, the VFNS predictions deviate rather little from the FFNS ones, typically by about 10\%, which is within the margins of renormalization and factorization scale uncertainties and ambiguities related to presently available parton distributions. Within the present intrinsic theoretical uncertainties, we can therefore rely on our uniquely generated NNLO VFNS parton distribution functions where, moreover, the required NNLO cross sections for massless initial-state partons are, in contrast to the fully massive FFNS, available in the literature for a variety of important production processes.

\begin{figure}
\begin{center}
\ifpdf
\includegraphics[width=0.6\textwidth]{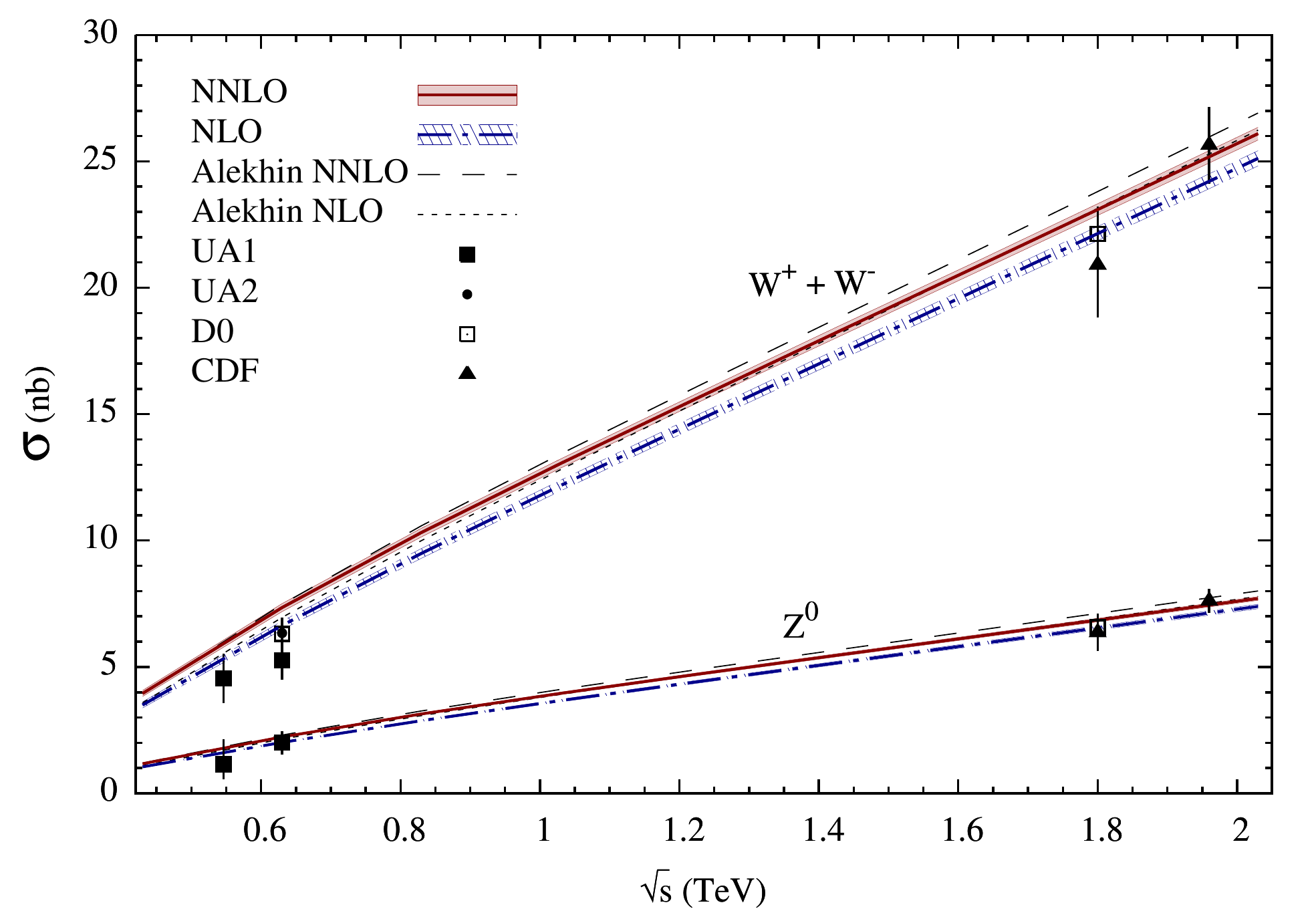}
\fi
\caption{Predictions for the total $W^++W^-$ and $Z^0$ production rates at $p\bar{p}$ colliders. The shaded band around our central results are due
to the $\pm 1\sigma$ PDF uncertainties. See \cite{JimenezDelgado:2009tv} for more details and references.}
\end{center}
\end{figure}

Our NNLO predictions \cite{JimenezDelgado:2009tv} for $\sigma(p\bar{p}\to W^{\pm}X)$ and $\sigma(p\bar{p}\to Z^0X)$ are compared with our NLO ones \cite{Gluck:2008gs} in Fig.\ 1, where also the predictions of Alekhin \cite{ref14,ref34} and some data points (see \cite{JimenezDelgado:2009tv} for the appropriate references) are shown for comparison. The vector boson production rates at NNLO are typically slightly {\em larger} (by more than $1\sigma$) than at NLO with a $K\equiv$ NNLO/NLO factor of $K^{W^+ +W^-}\!=\!1.04$ and $K^{Z^0}\!=\!1.06$ at Tevatron energies. This confirms the fast perturbative convergence at NNLO since the NLO/LO $K$-factor is of about 1.3 \cite{Gluck:2008gs}. The scale uncertainties of our NNLO predictions, due to $\frac{1}{2} M_V\leq \mu_F\leq 2M_V$, amount to less than 0.5\% at $\sqrt{s}\!=\!1.96$ TeV, i.e., is four times less than at NLO \cite{Gluck:2008gs}. Our results at $\sqrt{s}\!=\!1.96$ TeV are similar to the ones of MSTW \cite{Martin:2009bu} and about 4\% smaller than those of ABKM \cite{Alekhin:2009ni}.

Our NNLO expectations for $W^{\pm}$ and $Z^0$ production at the LHC at $\sqrt{s}\!=\!14$ TeV are:
\begin{eqnarray}
\sigma(pp\to W^+ +W^- +X) & = & 190.2 
         \pm 5.6_{\rm pdf}\,\,\,\,^{+1.6}_{-1.2}|_{\rm scale}\,\,{\rm nb}
\\
\sigma(pp\to Z^0 +X) & = & \,\,\,55.7 \pm
             1.5_{\rm pdf}\,\,\,\,^{+0.6}_{-0.3}|_{\rm scale}\,\,{\rm nb}\, .
\end{eqnarray}
Here the scale uncertainties amount to less than 1.7\%, i.e., are about half as large than the stated PDF uncertainties and than the scale uncertainties at NLO \cite{Gluck:2008gs}. These results are about 5\% smaller than the ones of MSTW \cite{Martin:2009bu} and about 10\% smaller than the obtained by ABKM \cite{Alekhin:2009ni}.  For comparison we note that within the FFNS the $W^+ +W^-$ production rate has been estimated \cite{Gluck:2008gs} to be about 192.7 nb at NLO with a total (PDF as well as scale) uncertainty of about 5\%. In general the NLO-VFNS prediction falls somewhat below that estimate but remains well within its total uncertainty of about 6\% \cite{Gluck:2008gs}. Due to the reduced scale ambiguity at NNLO and due to the slightly different NNLO estimates obtained by other groups, we conclude that the rates for gauge boson production at LHC energies can be rather confidently predicted with an accuracy of about 5\% irrespective of the factorization scheme.

\begin{figure}
\begin{center}
\ifpdf
\includegraphics[width=0.6\textwidth]{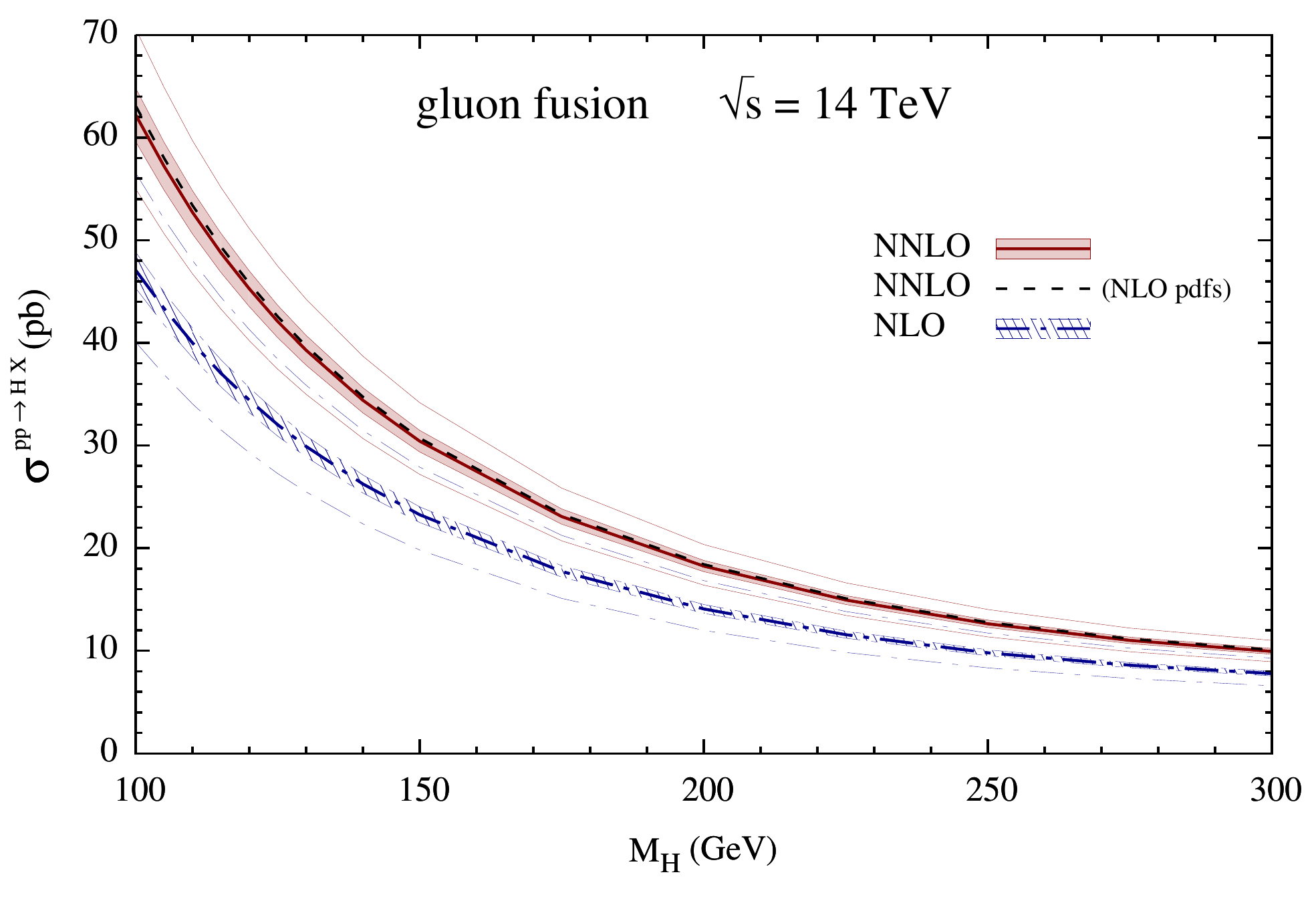}
\fi
\caption{Predictions for SM Higgs boson production at LHC via the dominant gluon-gluon fusion process. The shaded bands around the central values are due to the $\pm 1\sigma$ PDF uncertainties only, while the outer curves include both the PDF uncertainties and scale variations. See \cite{JimenezDelgado:2009tv} for more details.}
\end{center}
\end{figure}

We turn now to the hadronic production of the SM Higgs boson, where the dominant production mechanism proceeds via gluon-gluon fusion. Our NNLO and NLO results are shown in Fig.~2, where the shaded regions around the central predictions are due to the $\pm 1\sigma$ PDF uncertainties and the outer lines are obtained by varying the scale\footnote{In our calculations we always set $\mu_R\!=\!\mu_F$, as dictated by all presently available PDFs.} by a factor of 2 around its nominal value $\mu_F\!=\!M_H$ (see \cite{JimenezDelgado:2009tv} for a more explicit illustration of the scale ambiguities). Despite the fact that the NLO and NNLO total uncertainty bands overlap in Fig.\ 2, the predicted NNLO production rates are typically about 20\% larger than at NLO. The insensitivity of these predictions with respect to the appropriate choice of the PDFs is illustrated by the dashed curve, which has been obtained by using NNLO matrix elements and (inconsistently) NLO PDFs. Our central predictions in Fig.\ 2 are about 10\% smaller than the ones of MSTW \cite{Martin:2009bu}, and are 5-8\% smaller than those of ABKM \cite{Alekhin:2009ni} for $M_H$\raisebox{-0.1cm}{$\stackrel{<}{\sim}$}150~GeV, but agree with their predictions for larger Higgs masses \cite{Alekhin:2009ni}.

Higgs boson production at Tevatron have similar features than those shown in Fig.\ 2 but with much larger uncertainty bands; for instance, the uncertainties of our expectations at $\sqrt{s}\!=\!1.96$ TeV almost double at NNLO and NLO as compared to the ones at LHC \cite{JimenezDelgado:2009tv}. The rates obtained by ABKM \cite{Alekhin:2009ni} are 12-30\% smaller than our ones for $M_H\!=$100 - 200~GeV in this case. We conclude that SM Higgs boson production at LHC ($\sqrt{s}\!=\!14$ TeV) can be predicted with an accuracy of about 10\% at NNLO (with the total uncertainty being
almost twice as large at NLO), whereas the uncertainty almost doubles at Tevatron ($\sqrt{s}\!=\!1.96$ TeV).


\begin{thebibliography}{99}

  %\cite{Gluck:1994uf}
  \bibitem{Gluck:1994uf}
  M.~Gl\"uck, E.~Reya and A.~Vogt,
  %``Dynamical Parton Distributions Of The Proton And Small X Physics,''
  {\it Z.\ Phys.} {\bf C67} (1995) 433.
  %%CITATION = ZEPYA,C67,433;%%

  %\cite{Gluck:1998xa}
  \bibitem{Gluck:1998xa}
  M.~Gl\"uck, E.~Reya and A.~Vogt,
  %``Dynamical parton distributions revisited,''
  {\it Eur.\ Phys.\ J.} {\bf C5} (1998) 461.
  %[arXiv:hep-ph/9806404].
  %%CITATION = EPHJA,C5,461;%%

  %\cite{Gluck:2007ck}
  \bibitem{Gluck:2007ck}
  M.~Gl\"uck, P.~Jimenez-Delgado and E.~Reya,
  %``Dynamical parton distributions of the nucleon and very small-x physics,''
  {\it Eur.\ Phys. J.} {\bf C53} (2008) 355.
  %[arXiv:0709.0614 [hep-ph]].
  %%CITATION = EPHJA,C53,355;%%

  %\cite{JimenezDelgado:2008hf}
  \bibitem{JimenezDelgado:2008hf}
  P.~Jimenez-Delgado and E.~Reya,
  %``Dynamical NNLO parton distributions,''
  {\it Phys.\ Rev.} {\bf D79} (2009) 074023.
  %[arXiv:0810.4274 [hep-ph]].
  %%CITATION = PHRVA,D79,074023;%%

  %\cite{JimenezDelgado:2009tv}
  \bibitem{JimenezDelgado:2009tv}
  P.~Jimenez-Delgado and E.~Reya,
  %``Variable Flavor Number Parton Distributions and Weak Gauge and Higgs Boson Production at Hadron Colliders at NNLO of QCD,''
  {\it Phys.\ Rev.} {\bf D80} (2009) 114011.
  %[arXiv:0909.1711 [hep-ph]].
  %%CITATION = PHRVA,D80,114011;%%

  %\cite{Gluck:2008gs}
  \bibitem{Gluck:2008gs}
  M.~Gl\"uck, P.~Jimenez-Delgado, E.~Reya and C.~Schuck,
  %``On the role of heavy flavor parton distributions at high energy colliders,''
  {\it Phys.\ Lett.} {\bf B664} (2008) 133.
  %[arXiv:0801.3618 [hep-ph]].
  %%CITATION = PHLTA,B664,133;%%

  \bibitem{ref14}S.I.~Alekhin,
               {\it Phys. Rev.} {\bf D68} (2003) 014002.

  \bibitem{ref34}S.I.~Alekhin, 
               {\it JETP Lett.} {\bf 82} (2005) 628.

  %\cite{Martin:2009bu}
  \bibitem{Martin:2009bu}
  A.~D.~Martin, W.~J.~Stirling, R.~S.~Thorne and G.~Watt,
  %``Uncertainties on alpha_S in global PDF analyses and implications for
  %predicted hadronic cross sections,''
  {\it Eur.\ Phys.\ J.} {\bf C64} (2009) 653.
  %[arXiv:0905.3531 [hep-ph]].
  %%CITATION = EPHJA,C64,653;%%

  \bibitem{Alekhin:2009ni}
  S.~Alekhin, J.~Bl\"umlein, S.~Klein and S.~Moch,
  %``The 3-, 4-, and 5-flavor NNLO Parton from Deep-Inelastic-Scattering Data
  %and at Hadron Colliders,''
  {\it Phys.\ Rev.} {\bf D81} (2010) 014032.
  %[arXiv:0908.2766 [hep-ph]].
  %%CITATION = PHRVA,D81,014032;%%

\end{thebibliography}
\end{document}